\documentclass[reviewcopy]{elsart}
\usepackage{graphics}
\usepackage{amssymb}
\usepackage{graphicx}
\usepackage{dcolumn}
\usepackage{amsmath}

\def\1{\mathchoice{\rm 1\mskip-4.2mu l}{\rm 1\mskip-4.2mu l}{\rm
        1\mskip-4.6mu l}{\rm 1\mskip-5.2mu l}}
\begin{document}

\begin{frontmatter}
\title{Zero-field and Larmor spinor precessions\\ in a neutron polarimeter experiment}
\author{S. Sponar$^a$}, \author{J. Klepp$^a$}, \author{G.
Badurek$^a$}, and
\author{Y. Hasegawa$^{a,b}$\corauthref{YHasegawa}}\ead{Hasegawa@ati.ac.at}
\address{$^a$Atominstitut der {\"{O}}sterreichischen
Universit{\"{a}}ten, Stadionallee 2, A-1020 Vienna, Austria\\
$^b$PRESTO, Japan Science and Technology Agency (JST), Kawaguchi,
Saitama 332-0012, Japan}

\corauth[YHasegawa]{Corresponding author Phone +43-1-58801-14190;
Fax +43-1-58801-14199.}

\begin{abstract}
We present a neutron polarimetric experiment where two kinds of
spinor precessions are observed: one is induced by different total
energy of neutrons (zero-field precession) and the other is
induced by a stationary guide field (Larmor precession). A
characteristic of the former is the dependence of the
energy-difference, which is in practice tuned by the frequency of
the interacting oscillating magnetic field
$\omega_{\textrm{\scriptsize{R}}}$. In contrast the latter
completely depends on the strength of the guide field, namely
Larmor frequency $\omega_{\textrm{\scriptsize{L}}}$. Our
neutron-polarimetric experiment exhibits individual tuning as well
as specific properties of each spinor precession, which assures
the use of both spin precessions for multi-entangled spinor
manipulation.

\end{abstract}



\end{frontmatter}

In the last years great effort has been devoted to the development
of quantum information and communication technology \cite{00NC}.
Among them creation and manipulation of entanglement is a key
issue and experimental realizations with the use of photons,
atoms, ions, nuclear-magnet, superconducting qubits etc. have been
reported \cite{06HR}. There, the entanglement between spatially
separated systems are mainly used to exploit non-local property of
quantum systems. In contrast, investigations of two-qubit
entanglement with neutrons, one of the most useful quantum system
to be utilized for studies of quantum mechanical phenomena
\cite{00RW}, have been performed with the use of their degrees of
freedom, namely spin and path, in a single particle: a violation
of a Bell-like inequality \cite{03HLBBR}, contextual nature of
quantum theory \cite{06HLBBR}, and a full tomographic analysis of
Bell-states \cite{07HLBFKR} were carried out.

For further development towards multi-entanglement, it is
inevitable to establish manipulation of other degrees of freedom.
It is know that the total energy of neutrons can be manipulated
with the use of interactions between a neutron's magnetic moment
and a time-dependent oscillating magnetic field
\cite{81ABR,83BRS,93S}. Recently, we have accomplished an
experiment to exhibit a coherent energy manipulation of neutrons
with the use of radio-frequency (RF) oscillating magnetic field
\cite{08SKLFBHR}, enabling neutrons to be applied for
investigations of multi-entanglement. Thus, we find it significant
to demonstrate explicitly a phase manipulation for different
energy eigenstates which will be utilized in forthcoming
experiments such as a demonstration of triple-entanglment with a
GHZ-state \cite{89GHZ}. In the neutron scattering community, a
spin precession, alternatively a phase shift, due to energy
difference is known as zero-field precession \cite{94GGK,01KGLMR}
and utilized for zero-field spin-echo spectroscopy
\cite{87GG1,87GG2,02Y,04GKR}.

When a neutron is exposed to a stationary magnetic field, the motion of its spin vector is thoroughly described by the Bloch-equation exhibiting
Larmor precession:

\begin{equation}
\frac{d\vec S}{dt}=\gamma \vec S \times \vec B
\end{equation}

where $\gamma$ is the gyromagnetic ratio given by $\mu/\lvert S
\lvert$, $\mu$ and ${\vec S}$ are magnetic moment and the
neutron's spin, respectively. This is the equation of motion of a
classical magnetic dipole in a magnetic field, which shows the
precession of the spin vector ${\vec S}$ about the magnetic field
$\vec B$ with the Larmor frequency
$\omega_{\textrm{\scriptsize{L}}} = \left| {{{2\mu B}
\mathord{\left/ {\vphantom {{2\mu B} \hbar }} \right.
\kern-\nulldelimiterspace} \hbar }} \right|$. It is worth noting
here that the Larmor precession angle (rotation angle) is obtained
with a frequency $\omega_{\textrm{\scriptsize{L}}}$ and the
propagation time T as $\omega_{\textrm{\scriptsize{L}}} \rm{T}$,
solely depends on the strength of the applied magnetic field. In
practical experiments, the Larmor precession is utilized by a
variety of DC spin rotators.

In contrast, when a neutron interacts with a time-dependent,
rather oscillating, magnetic field, photon exchange occurs and the
total energy of neutrons is shifted. In particular when a
so-called resonant spin-flipper is activated, the neutron emits
(or absorbs) a photon of energy
$\hbar\omega_{\textrm{\scriptsize{R}}}$, thus the total energy of
the neutron decreases (or increases) by
$\hbar\omega_{\textrm{\scriptsize{R}}}$ during the interaction
\cite{93S}. As a consequence, an additional phase, a zero-field
phase, $\omega_{\textrm{\scriptsize{R}}} \rm{T}$ is accumulated
during the propagation afterwards. This phase shift, alternatively
a spin precession, due to energy shift emerges even without a
guide field, thus it is referred to a zero-field precession in
literature \cite{94GGK,01KGLMR}. It should be emphasized that this
zero-field phase is independent of the applied guide field, namely
the Larmor phase $\omega_{\textrm{\scriptsize{L}}} \rm{T}$, and
purely depends on the frequency of the spin-flipper. In the
following neutron polarimeter experiments, we demonstrate how one
can tune a zero-field phase $\omega_{\textrm{\scriptsize{R}}}
\rm{T}$ and a Larmor phase $\omega_{\textrm{\scriptsize{L}}}
\rm{T}$ independently.

\begin{figure}[!htbp]
          \begin{center}
            \scalebox{0.3}{\includegraphics{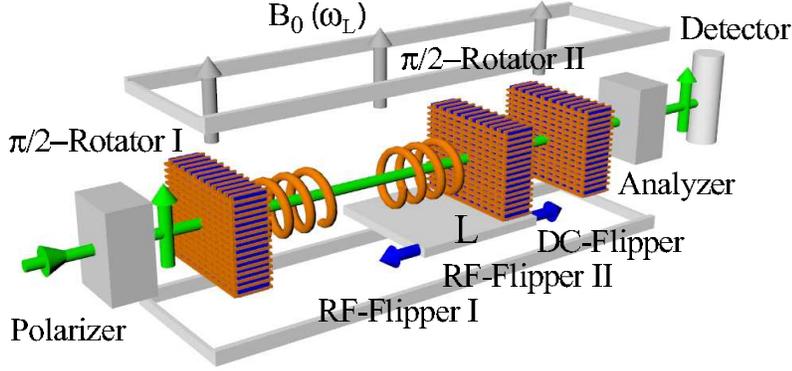}}
          \end{center}
          \caption{Experimental setup for demonstrations of zero-field and
          Larmor spin precessions. From a polarized incident beam the first $\pi/2$
          spin-rotator generates a superposition of $\lvert\pm z\rangle$.
          The first RF flipper induces the energy shift $\hbar \omega_R$ for
          each flipped spin eigenstate $\lvert\mp z\rangle$, which leads to the zero-field
          spin precession until the second RF flipper where the energy difference is compensated.
          The guide field $B_0$, permeated along
          the beam path, leads to the Larmor spin precession. The second
          $\pi/2$ spin-rotator together with the analyzer enables to exhibit
          the intensity modulation due to the spin precessions.
          The translation of the DC and the RF flippers manipulate individually the two phases
          when the RF flipper is turned on and off. }
            \label{Setup}
\end{figure}

The experiments were carried out at the neutron polarimetry
facility at the 250-kW TRIGA research reactor of the Atomic
Institute of the Austrian Universities in Vienna
\cite{97HMMB,99HB,00WBRBS,05JSHJB}. A schematic view of the
experimental setup is shown in Fig.1. The incident neutron beam is
monochromatized (with a mean wave length of $\lambda=1.99 \mbox{
\AA}$ which corresponds to a velocity of 1990 m/s) by the use of a
pyrolitic graphite crystal and polarized (average degree of
polarization over $98\%$) by reflection from a bent Co-Ti
supermirror array. The diameter of the beam is confined to about
4mm in diameter by a Cd diaphragm. The initial polarization vector
$\vec P_\textrm{i}$ is perpendicular to the beam trajectory and
defines the $+\hat{\mathbf z}$-direction. Another suppermirror
array is used to analyze the final polarization $\vec
P_\textrm{f}$. Depolarization of the neutron beam is minimized by
applying a guide field along the beam trajectory wherever
necessary: this guide field $B_0$ determines the Larmor frequency
$\omega_{\textrm{\scriptsize{L}}}$.

In a neutron polarimeter setup described above, after going
through a $\pi/2$ spin-rotator, an incident polarization is
interpreted as superposition of two orthogonal spin states
$\lvert\pm z\rangle$ and different phase shifts of these two
states under further spinor manipulations result in a polarization
change of the emerging neutron beam, i.e., we take a neutron
polarization-interference scheme. Between the two $\pi/2$
spin-rotators, three spin-flippers are placed in the beam: the
first two are RF spin-flippers followed by a DC spin-flipper. In
addition, one of the RF and the DC flippers are mounted together
on a single translator, allowing to tune the propagation time of
neutrons between the RF spin-flippers. The polarization vector
after passing through all spin-flippers is expected to lie in the
xy-plane, resulting from the interference between the states
$\lvert\pm z\rangle$. A spin analyzer together with the second
$\pi/2$ spin-rotator is used to resolve the different phase shifts
accumulated through spin-flips.

\begin{figure}[!htbp]
          \begin{center}
            \scalebox{0.25}{\includegraphics{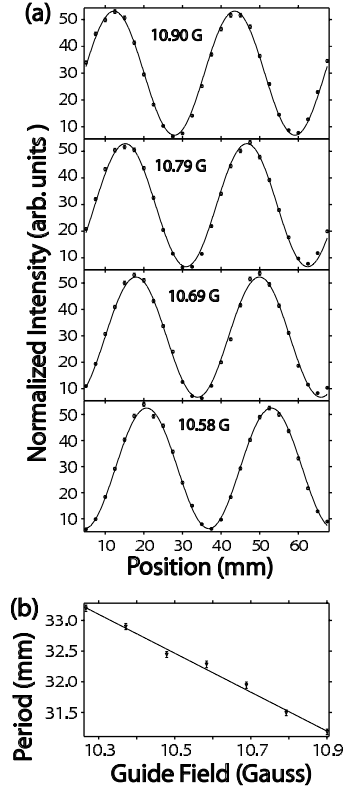}}
          \end{center}
          \caption{
          (a) Typical intensity oscillations with least square fits
          for the Larmor spin precession. Curves are shown
          with the guide field $B_0=$10.90, 10.79, 10.69,
          10.58G.(b) Dependence of the period on the strength
          of the guide field depicted for seven values of the guide field. A clear linear dependence agrees well
          with the theoretical predictions (see, Eq. (2)).
          }
            \label{Larmor1}
\end{figure}

The first experiment shows a pure Larmor precession: both RF
flippers are turned off and only the DC flipper is in operation.
In this case, the superposed states $\lvert\pm z\rangle$ (and the
flipped states $|\mp z\rangle$ later) simply propagate within the
guide field $B_0$. These states $|+z\rangle$ and $|-z\rangle$ are
the eigenstates of the guide field: no additional spinor rotation
occurs. Then, each state merely obtains a Larmor phase $\alpha$
due to the guide field, which is given by

\begin{eqnarray}
\alpha &=& {\omega }_\textrm{\scriptsize{L}} (\rm{T_1}-\rm{T_2}) \nonumber \\ &=&
{\omega }_\textrm{\scriptsize{L}} ((\rm{T_1^0}+\Delta T)-(\rm{T_2^0}-\Delta T)) \nonumber \\
&=&\alpha_0+2{\omega }_\textrm{\scriptsize{L}} \Delta \rm{T},
\end{eqnarray}

with $\alpha_0=\omega_\textrm{\scriptsize{L}}
(\rm{T_1^0}-\rm{T_2^0}+\rm{T_3^0})$. $\rm{T_1}$, $\rm{T_2}$,
$\rm{T_j^0}$ and $\Delta \rm{T}$ are the propagation time before
and after the DC flipper, the propagation time at the initial
position, and the time shift for $\rm{T_1}$, $\rm{T_2}$ by
shifting the position of the DC flipper. After the beam passes the
$\pi/2$ spin-rotator and the analyzer, clear sinusoidal intensity
oscillations of high contrast were obtained. Typical experimental
data with least square fits are shown in Fig.2(a): the strength of
the guide field was varied at $B_0=$10.90, 10.79, 10.69, 10.58G.
The shift of the oscillations is due to
$\omega_\textrm{\scriptsize{L}}\rm{T}$ (see, $\alpha_0$ in
Eq.(2)). In addition, the dependence of the period of the
oscillations on the strength of the guide field is plotted in
Fig.2(b). A linear dependence, of which inclination is in good
agreement with the theoretically predicted value, is seen, which
confirms the precession angle given by Eq.(2).

\begin{figure}[!htbp]
           \begin{center}
             \scalebox{0.25}{\includegraphics{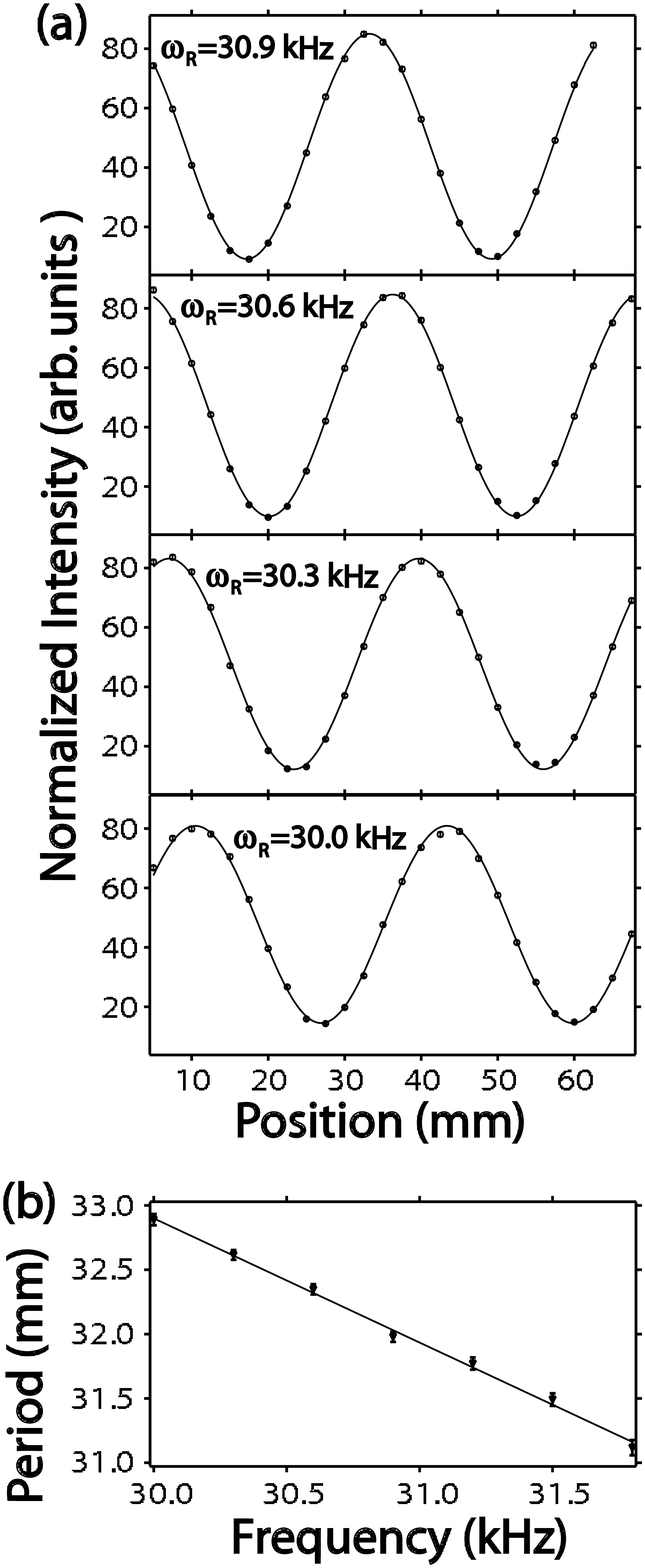}}
          \end{center}
          \caption{
          (a) Typical intensity oscillations with least square fits
          for the zero-field spin precession. Curves are
          shown at the RF flipper frequencies $\omega_R=$
          30.9, 30.6, 30.3, 30.0 kHz.
          (b) Dependence of the period on the frequency depicted for seven values of the frequency.
          A clear linear dependence agrees well
          with the theoretical predictions (see, Eq. (3)).
          }
            \label{ZF1}
\end{figure}

Next, characteristics of the zero-field precession are considered:
both RF flippers are turned on. In this case, the superposed
states $\lvert\pm z\rangle$ is once flipped to $\lvert\mp
z\rangle$, and finally again flipped back to $\lvert\pm z\rangle$.
Then, the spin precession angle $\alpha'$ is expected to be given
by the propagation time, $\rm{T_1}$, $\rm{T_2}$, and $\rm{T_3}$
after each spin flipper by

\begin{eqnarray}
\alpha' &=& {\omega}_\textrm{\scriptsize{L}} (\rm{T_1}-\rm{T_2}+\rm{T_3})+
{\omega}_\textrm{\scriptsize{R}} \rm{T_1} \nonumber \\
&=& {\omega }_\textrm{\scriptsize{L}} \{(\rm{T_1^0}+\Delta T)-\rm{T_2^0}+(\rm{T_3^0}-\Delta T)\} +{\omega}_\textrm{\scriptsize{R}}
(\rm{T_1^0}+\Delta T) \nonumber \\
&=& \alpha'_0+ {\omega }_\textrm{\scriptsize{R}} \Delta \rm{T},
\end{eqnarray}
with $\alpha'_0=\omega_\textrm{\scriptsize{L}}
(\rm{T_1^0}-\rm{T_2^0}+\rm{T_3^0})+\omega_\textrm{\scriptsize{R}}
\rm{T_1^0}$. In this configuration, no Larmor precession is
expected to be induced by the change of $\Delta \rm{T}$, since a
positive and a negative change of $\rm{T_1}$ and $\rm{T_3}$
completely compensate each other by shifting the position of the
DC flipper and the second RF flipper. In order to prove the
frequency dependence of the precession, the frequency of the RF
flippers were varied with keeping the strength of the guide field
by $B_0=$10.59G. Typical intensity modulations are shown in
Fig.3(a) at the RF flipper frequencies of
$\omega_{\textrm{\scriptsize{R}}}$=30.9, 30.6, 30.3, and 30.0kHz.
Clear sinusoidal oscillations of high contrast were again
obtained.  A slight reduction of the amplitude of the obtained
oscillations is solely due to the detuning of the frequency
$\omega_{\textrm{\scriptsize{R}}}$ in the vicinity of the
resonance for the flip-mode. And the shift of the oscillations
arises from $\omega_\textrm{\scriptsize{R}}\rm{T}$ (see,
$\alpha'_0$ in eq.(3)). In addition, a dependence of the period on
the frequency is plotted in Fig.3(b). A linear dependence is seen
and its inclination is in good agreement with the theory, which
confirms the precession angle given by Eq.(3).

\begin{figure}[!htbp]
           \begin{center}
              \scalebox{0.25}{\includegraphics{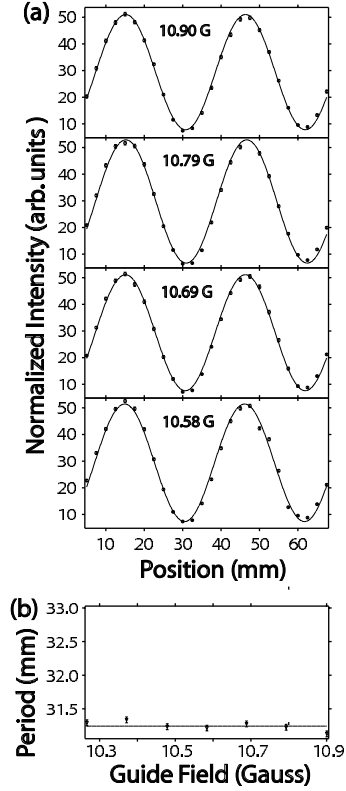}}
           \end{center}
          \caption{
          (a) Typical intensity oscillations with least square fits
          for the zero-field precession. Curves are
          shown with the guide field $B_0=$10.90, 10.79, 10.69, 10.58G.
          All curves are identical as expected.
          (b) The period is plotted versus the strength of the guide
          field for seven values. It is constant and
          independent of the strength of the guide field.
          }
            \label{ZF2}
\end{figure}

Note that the setup is constructed in a way that no spin rotation
due to the Larmor precession will occur, since the Larmor
precessions before and after the second RF flipper, i.e., $\Delta
\rm{T_1}$ and $\Delta \rm{T_3}$, are compensated each other with
all flippers turned on (see Eq.(3)). This independence of the
Larmor precession can also be accessed in our experimental setup.
In particular, the frequency of the RF flippers were tuned at the
resonance $\omega_0=$31.8kHz and the strength of the guide field
was varied at $B_0=$10.90, 10.79, 10.69, 10.58G. Typical
experimental data with least square fits are shown Fig.4(a):
identical sinusoidal oscillations are obtained, confirming no spin
rotation due to Larmor precessions. In addition, the period is
plotted versus the strength of the guide field in Fig.4(b).
Independent behavior of the period from the field strength is seen
as expected by Eq.(3).

The results are in good agreement with theoretically predicted
behavior: the frequency of the oscillating fields affects only on
the zero-field precession while the strength of the field does on
the Larmor precession. The tactics, that the RF and the DC
flippers are displaced by the same amount, allows individual
control of both precessions. This is a great advantage for the
separate (phase) manipulation of the two-spaces, i.e., spin and
energy spaces, in future neutron optical experiments. We are now
proceeding further neutron interferometric and polarimetric
experiments, where multi-entanglement in a single neutron system
will be investigated, in particular, with the use of entanglement
between degrees of freedoms such as path, spin, and energy.

In summary, we have presented experiments with a neutron
polarimeter where two spin-precessions are manipulated to show
their characteristics. The zero-field precession is independent of
the strength of the guide field and purely depends on the
frequency of the RF spin-flippers, where an energy shift occurs.
In contrast, the Larmor precession solely depends on the strength
of the applied magnetic field. In addition, we have exhibited a
method to manipulate individually the two phases, resulting from
each spin precession. The method used here will be utilized in
further neutron optical experiments dealing with
multi-entanglement in a single particle system.

This work has been partly supported by the Austrian Science
Foundation, FWF (P17803-N02). Y.H. would like to thank the Japan
Science and Technology Agency (JST) for financial support.


\end{document}